# Discrete Surface Modeling Based on Google Earth: A Case Study

Gang Mei, John C.Tipper
Institut für Geowissenschaften – Geologie
Albert-Ludwigs-Universität Freiburg
Freiburg im Breisgau, Germany
{gang.mei, john.tipper}@geologie.uni-freiburg.de

Nengxiong Xu
School of Engineering and Technology
China University of Geosciences (Beijing)
Beijing, China
xunengxiong@yahoo.com.cn

*Abstract*— **Google Earth (GE) has become a powerful tool for geological, geophysical and geographical modeling; yet GE can be accepted to acquire elevation data of terrain. In this paper, we present a real study case of building the discrete surface model (DSM) at Haut-Barr Castle in France based on the elevation data of terrain points extracted from GE using the COM API. We first locate the position of Haut-Barr Castle and determine the region of the study area, then extract elevation data of terrain at Haut-Barr, and thirdly create a planar triangular mesh that covers the study area and finally generate the desired DSM by calculating the elevation of vertices in the planar mesh via interpolating with Universal Kriging (UK) and Inverse Distance Weighting (IDW). The generated DSM can reflect the features of the ground surface at Haut-Barr well, and can be used for constructing the Sealed Engineering Geological Model (SEGM) in further step.**

*Keywords— Google Earth; Geological modeling; Elevation*

## I. INTRODUCTION

Google Earth (GE) achieved many but not all the elements of the first version of Digital Earth created by Gore in 1998 [1], and has become a powerful tool for geological, geophysical and geographical modeling. In [2], developments and scientific applications of GE were reviewed; the limitation and merit for Earth science studies at the global scale were also highlighted.

Keyhole Markup Language (KML) has become a standard form of representations of scientific data and an assortment of technical experiments, which can be used in virtual globes such as GE and the NASA World Wind. In [3], methods for creating KML files were summarized, and a guide for selecting tools to author KML for use with scientific data sets was introduced.

In order to understand more intuitively than conventional map symbols in 3D environment of a virtual globe, Blenkinsop [4] presented an Excel workbook S2K for converting field data into a KML document which can be used as a new method for visualizing the structural data.

Similar in [5], a visualizing package composed of the KML generator and user interface was presented for multidisciplinary geoscience data. The so-called package ''KML generators'' can convert the datasets of different fields of geosciences into KML files that would be used for visualization.

A Web-based and scalable visualization system based on the TCP/IP protocol and various data sharing open sources for climate research using Google Earth is developed in [6].

Chien and Tan [7] adopt GE as a tool in 2-D hydrodynamic modeling. They first draw polygons on the GE screen and save these features into a KML file, and convert the resulting output text file from the hydrodynamic simulation back to KML.

De Paor and Whitmeyer [8] developed some KML codes to render geological maps and link associated COLLADA models to represent data such as structural orientations, cross-sections, and geophysical moment tensor solutions by using the GE API with multiple Javascript controls.

In this paper, we present a useful approach for building the discrete surface model for the surface at the selected study area, Haut-Barr, based on the elevation data extracted from GE.

This paper is organized as follows. In Sect.2, the method of building discrete surface models is described in details. Sect.3 gives the implementation of a real study-case and shows how to create the DSM of the selected study area. Finally, in Sect.4 we summarize our work in this paper.

## II. THE METHODOLOGY

The 3D geological modeling technique is a powerful tool to depict geological and engineering realities. Combining it with numerical simulation can effectively improve the reliability of calculating results and the efficiency of modeling.

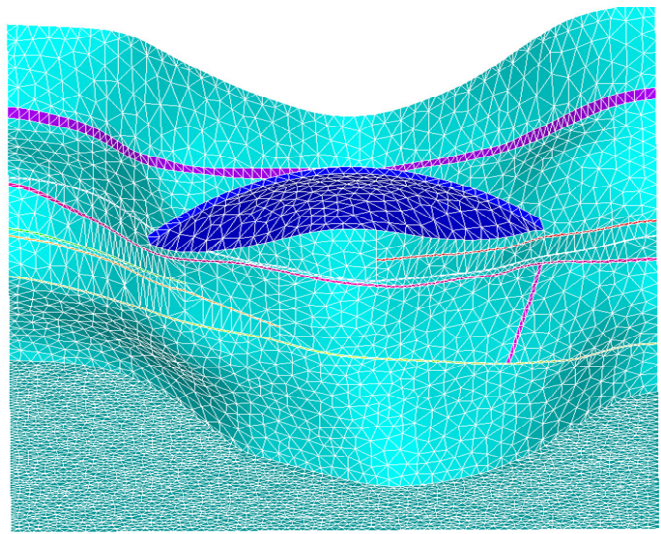

Figure 1.  An example of 3D SEGM from [10]

This research was supported by the Natural Science Foundation of China (Grant Numbers 40602037 and 40872183) and Fundamental Research Funds for the Central Universities of China.

When construct the 3D geological models, especially the Sealed Engineering Geological Model (SEGM) [9] that is quite suitable for numerical simulation (Figure 1), the interpretation or representation of the earth surface / terrain is critical.

The Discrete Surface Model (DSM) is a form for reflecting ground surface using a discrete mesh / grid that consists of a set of nodes (points) and elements (polygons).

In this paper, we put our emphasis mainly on constructing a real and accurate DSM based on the elevation data of terrain points extracted from GE, rather than drawing contour maps or creating Digital Elevation Model (DEM).

*A. Overview*

Our procedure of constructing the DSM can be summarized mainly into four steps:

Step 1: Extract the elevation date of terrain points from GE using the GE COM API.

Step 2: Transform the extracted coordinates of points from geographic coordinates system WGS-84 to UTM.

Step 3: Create and then smooth planar triangular mesh that covers the region of study area.

Step 4: Interpolating with Kriging and IDW to obtain the elevation of each vertex of the planar mesh to form DSM.

*B. Extracting elevation data from GE*

In order to allow the GE users to develop their applications, GE has provided some application programming interface (API) including Google Map and Google Earth API. The GE COM API provides several useful interfaces that allow users to call and use them to manipulate the GE interactively such as saving the screenshot, creating geometries, loading KMZ / KML files and extracting elevation of a point displayed on the screen.

The GE COM API consists of 11 powerful classes; and one of the most important is the IApplicatonGE which can be used to realize most functionalities of the GE client by catching the program handle of the process of GE.

There is a quite useful function in the class IApplicationGE, GetPointOnTerrainFromScreenCoords, which is used to locate a point on the GE terrain according a position (double screen_x, double screen_y) on the screen.

The acquired point via the above function is an object of a class IPointOnTerrainGE that has the properties of coordinates represented as Latitude, longitude and Altitude.

Obviously, before we try to acquire a point on the terrain, it is necessary to make it displayed on the screen of GE. This can be realized using another function IApplicationGE::SetCamera which is designed for controlling the setting of views.

In order to extract the elevation of terrain points from GE, we first need to control the GE process by obtaining its handle, and then change the views using SetCamera to make the target region or points be displayed on the screen, and thirdly call the key function GetPointOnTerrainFromScreenCoords to receive a terrain point with its Altitude.

In summary, the procedure of extracting the elevation data from GE includes: determining the region to be extracted from, setting the parameters for the camera, specifying the number of points to be extracted by a grid with given rows and columns, and finally extracting the elevation of points row by row

*C. Converting to UTM*

Two common coordinate systems, geographic coordinate system WGS-84 and Universal Transverse Mercator (UTM), are provided in GE. The coordinates of points extracted from GE is formatted in geographic coordinate system WGS-84.

These coordinates represented by Latitude, Longitude and Altitude are not suitable for building accurate DSM. Thus, it is necessary to convert the coordinates from WGS-84 to UTM.

In [11], an effective and easily-used spreadsheet is provided for converting above two kinds of coordinates system into each other; and we directly adopt this Excel workbook to transform the extracted points from WGS-84 to UTM.

*D. Gernerating and smoothing planar mesh*

Delaunay Triangulation (DT) is maybe the most commonly accepted technique for generating triangular meshes in 2D and tetrahedral meshes in 3D [12, 13]. In this paper, we first locate the four corners of the selected region, and then create a planar triangular mesh in the above rectangular area.

In order to improve qualities of the planar triangular mesh, Laplacian smoothing [14] is accepted after creating the original mesh. Noticeably, the mesh smoothing for the planar meshes in this step may not be meaningful in some situations for that the elements of the triangular mesh will alter after interpolating.

The motivation of conducting mesh smoothing at this time is to try to avoid bad triangles before interpolated and to reduce even worse triangles after interpolating.

*E. Interpolating*

Two interpolation algorithms widely used in geostatistics, Universal Kriging(UK) and Inverse Distance Weighting(IDW), are accepted to obtain the elevation of nodes in planar mesh.

*1) Universal Kriging*

Kriging was originally developed in geostatistics by South African mining engineer D.G.Krige [15]. The mathematics was further developed by Matheron [16] in 1963.

For a set of regionalized variables $Z(x)$, Kriging estimates the expected value $Z(x_0)$ at location $x_0$ where the observation is not available using a linear weighted sum of the known values $Z(x_1), Z(x_2), ..., Z(x_n)$ at locations $x_1, x_2, ..., x_n$, such that:

$$Z(x_0) = \sum_{i=1}^{n} \lambda_i \cdot Z(x_i), \text{ with } \sum_{i=1}^{n} \lambda_i = 1 \qquad (1)$$

In (1), $\lambda_i$ is the weights at the location $x_i$ which can be calculated according to the equations of UK.

UK can use either semivariograms or covariances; the mathematical form of the UK equations expressed with semivariograms is as follows:

$$\begin{cases} \sum_{j=1}^{n} \lambda_j \cdot \gamma_{ij}(x_i, x_j) + \sum_{l=0}^{k} \mu_l f_l(x_i) = \gamma_{i0}(x_i, x_0) & i = 1, 2, \cdots, n \\ \sum_{i=1}^{n} \lambda_i f_l(x_i) = f_l(x_0) & l = 0, 1, \cdots, k \end{cases} \quad (2)$$

In UK, the trend component is no longer considered being constant but usually assumed as:

$$m(x) = \sum_{l=0}^{k} \mu_l f_l(x) \quad (3)$$

If the degree $k = 1$, then the local trend is linear and (3) will then be $m(x) = \mu_0 + \mu_1 x + \mu_2 y$. Correspondingly, the equations of UK shown in (2) can be expressed as the matrix form (4).

$$\begin{bmatrix} \gamma_{11} & \gamma_{12} & \cdots & \gamma_{1n} & 1 & x_1 & y_1 \\ \gamma_{21} & \gamma_{22} & \cdots & \gamma_{2n} & 1 & x_2 & y_2 \\ \vdots & \vdots & \ddots & \vdots & \vdots & \vdots & \vdots \\ \gamma_{n1} & \gamma_{n2} & \cdots & \gamma_{nn} & 1 & x_n & y_n \\ 1 & 1 & \cdots & 1 & 0 & 0 & 0 \\ x_1 & x_2 & \cdots & x_n & 0 & 0 & 0 \\ y_1 & y_2 & \cdots & y_n & 0 & 0 & 0 \end{bmatrix} \times \begin{bmatrix} \lambda_1 \\ \lambda_2 \\ \vdots \\ \lambda_n \\ \mu_0 \\ \mu_1 \\ \mu_2 \end{bmatrix} = \begin{bmatrix} \gamma_{10} \\ \gamma_{20} \\ \vdots \\ \gamma_{n0} \\ 1 \\ x_0 \\ y_0 \end{bmatrix} \quad (4)$$

A key issue in UK is to calculate variograms. Experimental variogram are estimates of the theoretical ones. The variograms are calculated for a set of vector $h$ that is in fact the distance of any pair of observations.

Commonly used variogram theoretical models include: the spherical model (5), Gaussian model and exponential model.

$$\gamma(h) = \begin{cases} 0 & h = 0 \\ C_0 + C\left(\frac{3}{2} \cdot \frac{h}{a} - \frac{1}{2} \cdot \frac{h^3}{a^3}\right) & 0 < h \leq a \\ C_0 + C & h > a \end{cases} \quad (5)$$

$C_0$ (nugget): The height of the jump of the semivariogram at the discontinuity at the origin, which is the sum of geological microstructure and measurement error.

$C$ (sill): Limit of variogram tending to infinity lag distances; in some applications, it can be set to be equal to $C_0$.

$a$ (range): The distance which separates the correlated and the uncorrelated random variables; also, it is the distance at which the variogram reaches the sill.

*2) Inverse Distance Weighting*

Similar to Kriging, IDW [16] also obtain the expected values of unknown points (interpolated points) by weighting average of the values of known points (data points). The name given to this type of methods was motivated by the weighted average applied since it resorts to the inverse of the distance to each known point when calculate the weights. The difference between IDW and Kriging is that they calculate the weights variously.

A general form of finding an interpolated value $Z$ at a given point $x$ based on samples $Z_i = Z(x_i)$ for $i = 1, 2, \ldots, n$ using IDW is an interpolating function:

$$Z(x) = \sum_{i=1}^{n} \frac{\omega_i(x) z_i}{\sum_{j=1}^{n} \omega_j(x)}, \quad \omega_i(x) = \frac{1}{d(x, x_i)^p} \quad (6)$$

Above (6) is a simple IDW weighting function, as defined by Shepard [17], $x$ denotes an interpolated (arbitrary) point, $x_i$ is an interpolating (known) point, $d$ is a given distance (metric operator) from the known point $x_i$ to the unknown point $x$, $n$ is the total number of known points used in interpolation and $p$ is a positive real number, called the power parameter.

### III. CASE STUDY

*A. The study area*

The Haut-Barr, as shown in Figure 2, is a medieval castle, first built in 1100 on a rose-colored sandstone rock 460m above the valley of Zorn and the plain of Alsace in France. The rock bar made of three consecutive rocks, i.e., the Septentrional Rock, the Median Rock and the Southernmost Rock that is best known as Markenfels. Two of them are joined by a foot bridge named Devil's Bridge. Since the 12th century, the Haut-Barr Castle has been built by Rodolphe, bishop of Strasbourg.

A rectangular region covering the study area is located by 4 corners. The coordinates of these corners are listed in Table 1. The distances along Latitude and Longitude are both 12.96" in WGS-84, and about 300m and 400m in UTM, respectively.

TABLE I. COORDINATES OF 4 CORNERS OF THE STUDY AREA

| Point | WGS-84 | | UTM ( /m) | |
|---|---|---|---|---|
| | *Latitude* | *Longitude* | *X / Easting* | *Y / Northing* |
| 1 | N48°43'20.64" | E7°20'12.48" | 377676.932 | 5397932.106 |
| 2 | N48°43'20.64" | E7°20'25.44" | 377941.691 | 5397926.333 |
| 3 | N48°43'33.6" | E7°20'25.44" | 377950.404 | 5398326.487 |
| 4 | N48°43'33.6" | E7°20'12.48" | 377685.664 | 5398332.260 |

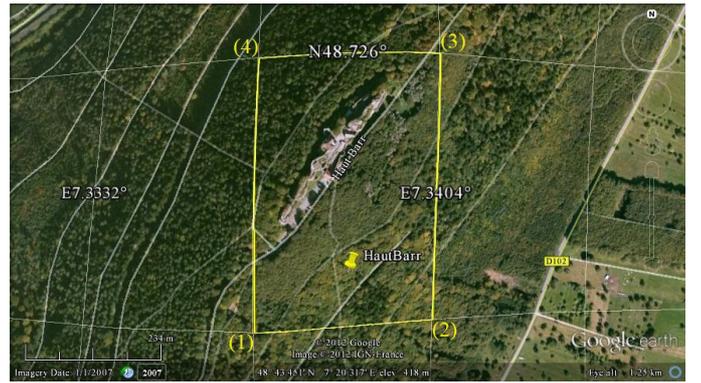

Figure 2. Location of the study area (Haut-Barr)

*B. Implementation of modeling*

*1) Extraction*

A small code named GetElevation has been developed for extracting elevation data from GE (see Appendix), which scans the points that assigned inside the study area row by row.

Noticeably, in order to obtain enough original points being extracted, an optical solution is to create a little bigger region when extract and then discard those points located outside the target region. 5000 points (by 100 rows and 50 columns) are created as the target to be extracted. Contour maps have been created using the extracted scatter points, as shown in Figure 3.

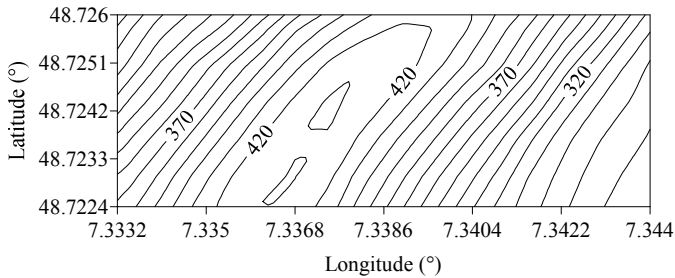

(a) Contour map of an extended region

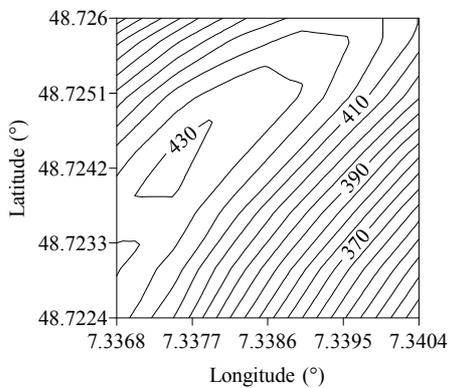

(b) Contour map of the study area

Figure 3.  Contour maps created according to the extracted elevation data.

*2) DSM*

After extracting elevations, later steps including generating and smoothing the planar triangular mesh and interpolating with Kriging and IDW have been implemented. The original surface at Haut-Barr was displayed in GE (Figure 4); and the generated DSMs are shown in Figure 5. The DSM interpolated by Kriging is a little smoother than that by IDW.

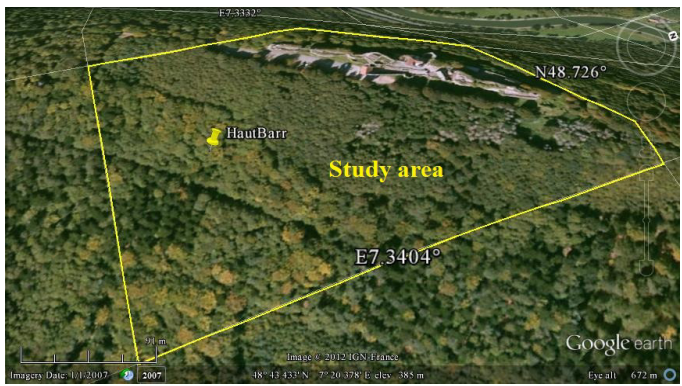

Figure 4.  Ground surface of the study area viewed in GE

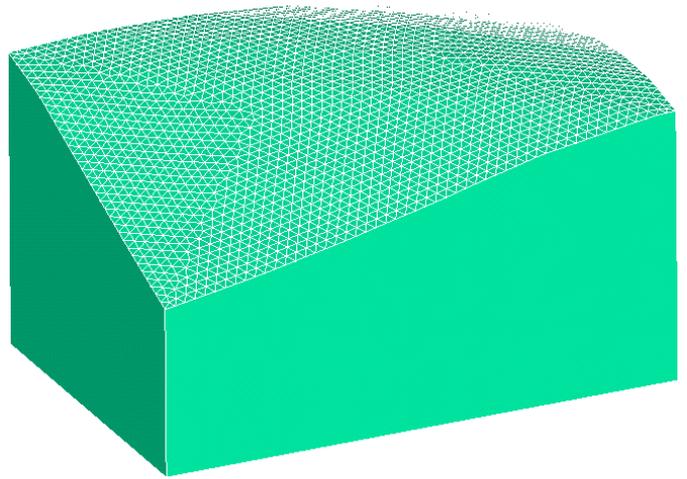

(a) DSM interpolated by Kriging

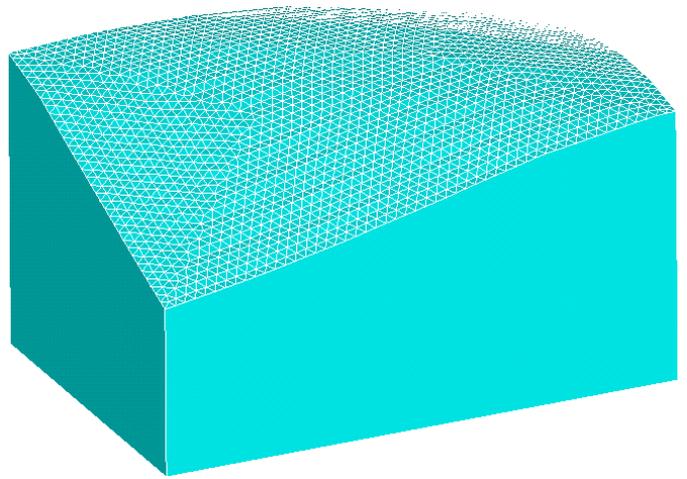

(b) DSM interpolated by IDW

Figure 5.  DSMs of the study area (Haut-Barr)

## IV. Summary

In this paper, we present a solution for building the discrete surface model (DSM) based on the elevation data of terrain points extracted from GE using the COM API, and give a real study case of constructing DSM for Haut-Barr Castle, France.

We first locate the position of Haut-Barr and determine the region of the study area, then extract the elevation data of the terrain of Haut-Barr, and thirdly create a planar triangular mesh covering the selected study area and finally obtain the DSM by obtaining elevation of vertices via interpolating with Universal Kriging (UK) and Inverse Distance Weighting (IDW).

The generated DSM can reflect the realities of the ground surface at Haut-Barr well, and can be used for constructing the Sealed Engineering Geological Model (SEGM) in further step.

Future works may focus on acquiring the elevation data for huge number of points in multithread or parallel to reduce time cost. Another problem is to improve the quality of DSM since that some triangles may badly shape due to extruding after interpolating by UK or IDW.

## APPENDIX

*A. C# code for extracting elevation data from GE*

```csharp
using EARTHLib;    // API
namespace GetElevation {
  class Program {
    static void Main(string[] args) {
      ApplicationGEClass
      GeApp = new ApplicationGEClass();    // GE
      GeApp.SetCameraParams(48.7242, 7.3386, 0,
      (AltitudeModeGE)1, 600, 0, 0, 5);    // Camera
      FileInfo f = new FileInfo("ElevGE.txt");
      StreamWriter w = f.CreateText();    // Save

      int Col = 100, Row = 50;
      double distCol = 2.0 / (double) Col;
      double distRow = 2.0 / (double) Row;
      double scrX, scrY;    // Coordinates

      IPointOnTerrainGE pt;
      for (int i = 0; i < Row; i++) {
        for (int j = 0; j < Col; j++) {
          scrX = -1 + (double) j * distCol;
          scrY = 1 - (double) i * distRow;
          pt = GeApp.GetPointOnTerrainFromScreenCoords
               (scrX, scrY);    // Point
          w.Write(pt.Latitude); w.Write(' ');
          w.Write(pt.Longitude); w.Write(' ');
          w.Write(pt.Altitude); w.WriteLine();
        }
      }
      w.Close();    // Finish saving
    }
  }
}
```

ACKNOWLEDGMENT

The corresponding author Gang Mei would like to thank Chun Liu at Universität Kassel for sharing several interesting and valuable ideas.